\documentclass[prb,twocolumn,floatfix,preprintnumbers,amsmath,amssymb,superscriptaddress]{revtex4}

\usepackage{graphicx}% Include figure files
\usepackage{amsmath}
\usepackage{dcolumn}% Align table columns on decimal point
\usepackage{bm}% bold math
\usepackage{color}
\usepackage{xspace}
\usepackage{float}
\usepackage[normalem]{ulem}

\begin{document}

\title{Spin pseudogap in the $S=\frac{1}{2}$ chain material Sr$_2$CuO$_3$ with impurities}

\author{G. Simutis}
\email{gsimutis@phys.ethz.ch}
\affiliation{Neutron Scattering and Magnetism, Laboratory for Solid State
Physics, ETH Z\"urich, Z\"urich, Switzerland}

\author{S. Gvasaliya}
\affiliation{Neutron Scattering and Magnetism, Laboratory for Solid State
Physics, ETH Z\"urich, Z\"urich, Switzerland}

\author{N. S. Beesetty}
\affiliation{Synthese, Proprietes et Modelisation des Materiaux, Universite Paris-Sud, 91405 Orsay cedex, France}

\author{T. Yoshida}
\affiliation{Institute for Solid State Physics, University of Tokyo, 5-1-5 Kashiwanoha, Kashiwa, Chiba 277-8581, Japan}

\author{J. Robert}
\affiliation{Laboratoire L{\'e}on Brillouin, CEA–CNRS, CEA-Saclay, F-91191 Gif-sur-Yvette, France}

\author{S. Petit}
\affiliation{Laboratoire L{\'e}on Brillouin, CEA–CNRS, CEA-Saclay, F-91191 Gif-sur-Yvette, France}

\author{A. I. Kolesnikov}
\affiliation{Chemical and Engineering Materials Division, Oak Ridge National Laboratory, Oak Ridge, TN 37831, USA}

\author{M. B. Stone}
\affiliation{Quantum Condensed Matter Division, Oak Ridge National Laboratory, Oak Ridge, TN 37831-6393, USA}

\author{F. Bourdarot}
\affiliation{Mod\'elisation et d’exploration de la mati\'ere , Univ. Grenoble Alpes et CEA, INAC, 17 rue des Martyrs, 38054 Grenoble, France}

\author{H. C. Walker}
\affiliation{ISIS Facility, Rutherford Appleton Laboratory, Chilton, Didcot, Oxon OX11 OQX, United Kingdom}

\author{D.T. Adroja}
\affiliation{ISIS Facility, Rutherford Appleton Laboratory, Chilton, Didcot, Oxon OX11 OQX, United Kingdom}

\author{O. Sobolev}
\affiliation{Forschungsneutronenquelle Heinz Maier-Leibnitz (FRM-II), TU München, D-85747 Garching, Germany}

\author{C. Hess}
\affiliation{Leibniz Institute for Solid State and Materials Research IFW Dresden, P.O. Box 270116, D-01171 Dresden, Germany}

\author{T. Masuda}
\affiliation{Institute for Solid State Physics, University of Tokyo, 5-1-5 Kashiwanoha, Kashiwa, Chiba 277-8581, Japan}

\author{A. Revcolevschi}
\affiliation{Synthese, Proprietes et Modelisation des Materiaux, Universite Paris-Sud, 91405 Orsay cedex, France}

\author{B. B\"uchner }
\affiliation{Leibniz Institute for Solid State and Materials Research IFW Dresden, P.O. Box 270116, D-01171 Dresden, Germany}

\author{A. Zheludev}
\affiliation{Neutron Scattering and Magnetism, Laboratory for Solid State
Physics, ETH Z\"urich, Z\"urich, Switzerland}

\date{\today}

\begin{abstract}
The low energy magnetic excitation spectrum of the Heisenberg antiferromagnetic $S = 1/2$ chain system Sr$_2$CuO$_3$ with Ni- and Ca-impurities is studied by neutron spectroscopy. In all cases, a defect-induced spectral pseudogap is observed and shown to scale proportionately to the number of scattering centers in the spin chains.
\end{abstract}

 \maketitle
\section{Introduction}

Defects play a special role in one dimension due to its intrinsic topology.\cite{Eggert2002b,Kane1992} This is especially pronounced in quantum magnets where both the ground states and the excitations are modified upon introduction of imperfections in magnetic lattices.\cite{Dasgupta1980,Affleck1987,Fisher1994,Damle2000,Motrunich2001,Eggert2002,Schmitteckert1998,Giamarchi1987} Recent experiments have demonstrated that when impurities are added to prototypical Heisenberg Antiferromagnetic (HAF) $S=1/2$ chain materials SrCuO$_2$ and Sr$_2$CuO$_3$, their magnetism is suppressed.\cite{Kojima2004, Hammerath2011, Simutis2013, Karmakar2015,Utz2015, Simutis2016b} Long range magnetic ordering arising from residual three-dimensional interactions becomes inhomogeneous and occurs at lower temperatures.\cite{Kojima2004, Simutis2016b} Additionally, the spin-lattice relaxation rate 1/T$_1$ measured by NMR experiments drops at low temperatures, suggesting a depletion of low energy excitations.\cite{Hammerath2011,Hammerath2014,Utz2015} Defects also affect the heat transport and compromise its ballistic nature.\cite{Hlubek2011,Mohan2014}

The influence of defects on the magnetic excitation spectrum can be directly probed using neutron scattering. We have previously employed this method to demonstrate the opening of a spin pseudogap in the  Heisenberg spin chain system SrCuO$_2$ with chain-breaking Ni$^{2+}$ impurities.\cite{Simutis2013} We proposed a simple chain-fragmentation model based on previous theoretical work,\cite{Eggert1992} and derived quantitative predictions for the excitation spectrum in the presence of defects. This model was able to account for the data in a wide temperature range, on an absolute scale and with no adjustable parameters. Nevertheless, a few questions remained to be answered: i) Does the size of the pseudogap indeed scale with the impurity concentration as predicted?  ii) What is the role of the double-chain structure of SrCuO$_2$?  iii) What effect would other types of impurities have?

\begin{figure} [h!]
\centering
\includegraphics[width={\columnwidth}]{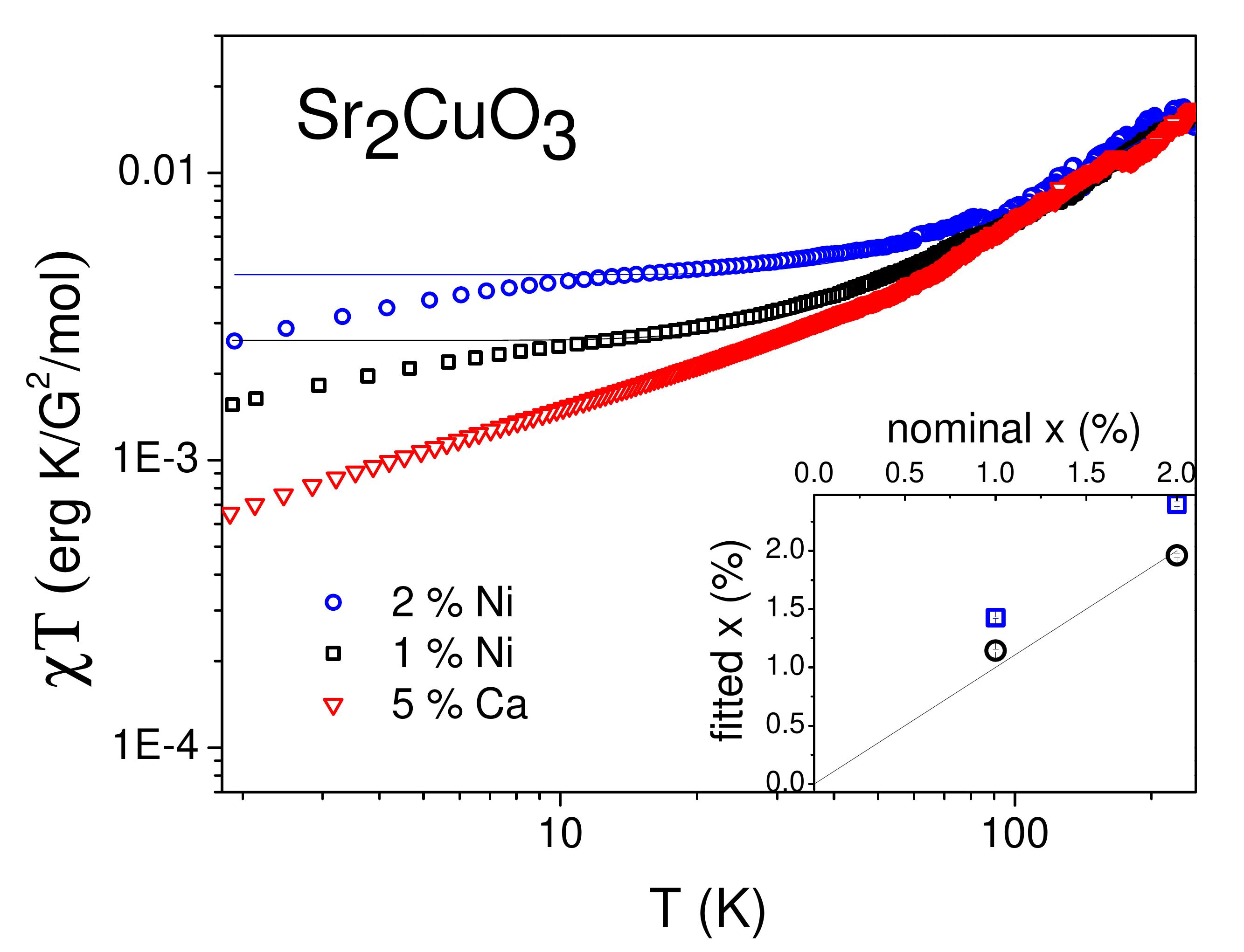}
\caption{Magnetic susceptibility multiplied by temperature plotted vs. temperature for the studied compounds (symbols). The solid lines are fits as described in the text. The extracted impurity concentration of Ni-substituted samples is shown in the inset for fit using all the data (black circles) and measurements only above 10K (blue squares). The solid black line corresponds to the number of impurities equal to the nominal concentration. In the case of Ca substitution, the magnetic response is much smaller as expected from substitution outside the chain. \label{fig:suscep}}
\end{figure}

The present paper aims to settle these outstanding issues.  Here we study a related spin chain compound, namely Sr$_2$CuO$_3$. It has the benefit of having a simpler single-chain structure, as opposed to the paired chains in SrCuO$_2$.\cite{Motoyama1996,Kojima1997,Walters2009} Two different types of impurities are investigated: $S = 1$ Ni$^{2+}$ ions that replace the $S=1/2$ Cu$^{2+}$ ions in the chain,\cite{Eggert1992,Kojima2004,Simutis2016b} and Ca$^{2+}$ ions that replace Sr$^{2+}$ and therefore affect the Cu$^{2+}$ chains only indirectly.\cite{Mohan2014,Hammerath2014}

\begin{figure}
\centering
\includegraphics[width={\columnwidth}]{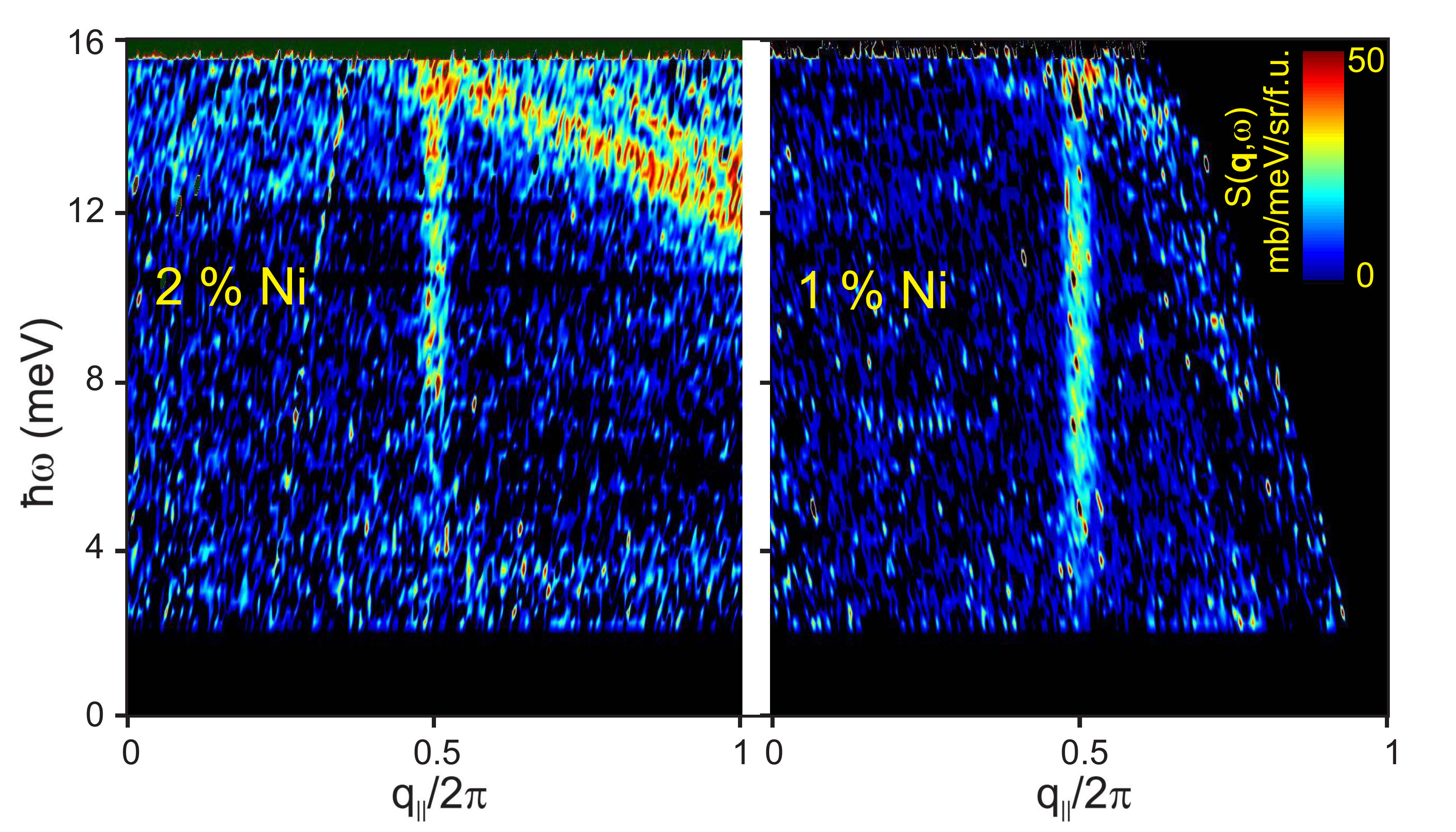}
\caption{Dynamical structure factor of the two Ni-substituted samples as measured by SEQUOIA time-of-flight neutron spectrometer. The vertical rods at the antiferromagnetic wavevector are the bottom of the highly dispersive two-spinon continuum. The suppression of low energy states give rise to a pseudogap which scales in energy with increasing number of broken links as described in the text. \label{fig:Ni_slice}}
\end{figure}

\section{Experiment}

Single crystals of Sr$_2$CuO$_3$ with Ni and Ca impurities were grown using floating zone furnaces as described in earlier reports.\cite{Mohan2014,Utz2015} They crystalize in an orthorhombic {\it Immm} space group, with lattice parameters of $a = 3.9089,$ $b =3.4940$ and $c=12.6910$ \AA $ $  for the pure compound.\cite{Ami1995} The spin chains formed by the Cu$^{2+}$ ions run along the crystallographic \textbf{b} direction. The exchange interaction between the spins has been estimated at $J=241(11)$ meV.\cite{Walters2009} While the chains are very well isolated, the three-dimensional ordering still takes place due to residual three-dimensional interactions. In the pure system the spins order at $T_N = 5.4$ K\cite{Kojima1997} and the ordering temperature is rapidly reduced with the introduction of impurities.\cite{Kojima2004} All of the measurements presented here were performed above the three-dimensional ordering temperatures. The measurement temperatures were much smaller than the temperature corresponding to the intrachain exchange energy.

In order to study the effects of disorder, diligent attention had to be given to controlling the level of impurities. In addition to carefully monitoring the ingredients in the growth procedure and performing energy dispersive X-ray analysis measurements, we have also measured magnetic susceptibility in order to estimate the extent of chain fragmentation. The small samples used for susceptibility measurements were cut from the same crystals that were used for neutron spectroscopy. The crystals were aligned with the \textbf{b} crystal axis parallel to the applied magnetic field. The data were taken with the vibrating sample magnetometer option of the Quantum Design PPMS instrument.

The samples studied with neutron spectroscopy were made of a few co-aligned crystals with the crystal \textbf{a} axis perpendicular to the scattering plane. The sample with 1 \% Ni impurities consisted of three co-aligned single crystals with a total mass of 8.6 g and a FWHM mosaic of 0.4$^\circ$ as measured at the 020 and 002 peaks. The sample with 2 \% Ni impurities consisted of two single crystals with a mass of 4.6 g and mosaic of  0.7$^\circ$.  The sample with 5 \% Ca impurities was made of two crystals with a total mass of 3.5 g and a mosaic of 0.3$^\circ$.

\begin{figure}
\centering
\includegraphics[width={\columnwidth}]{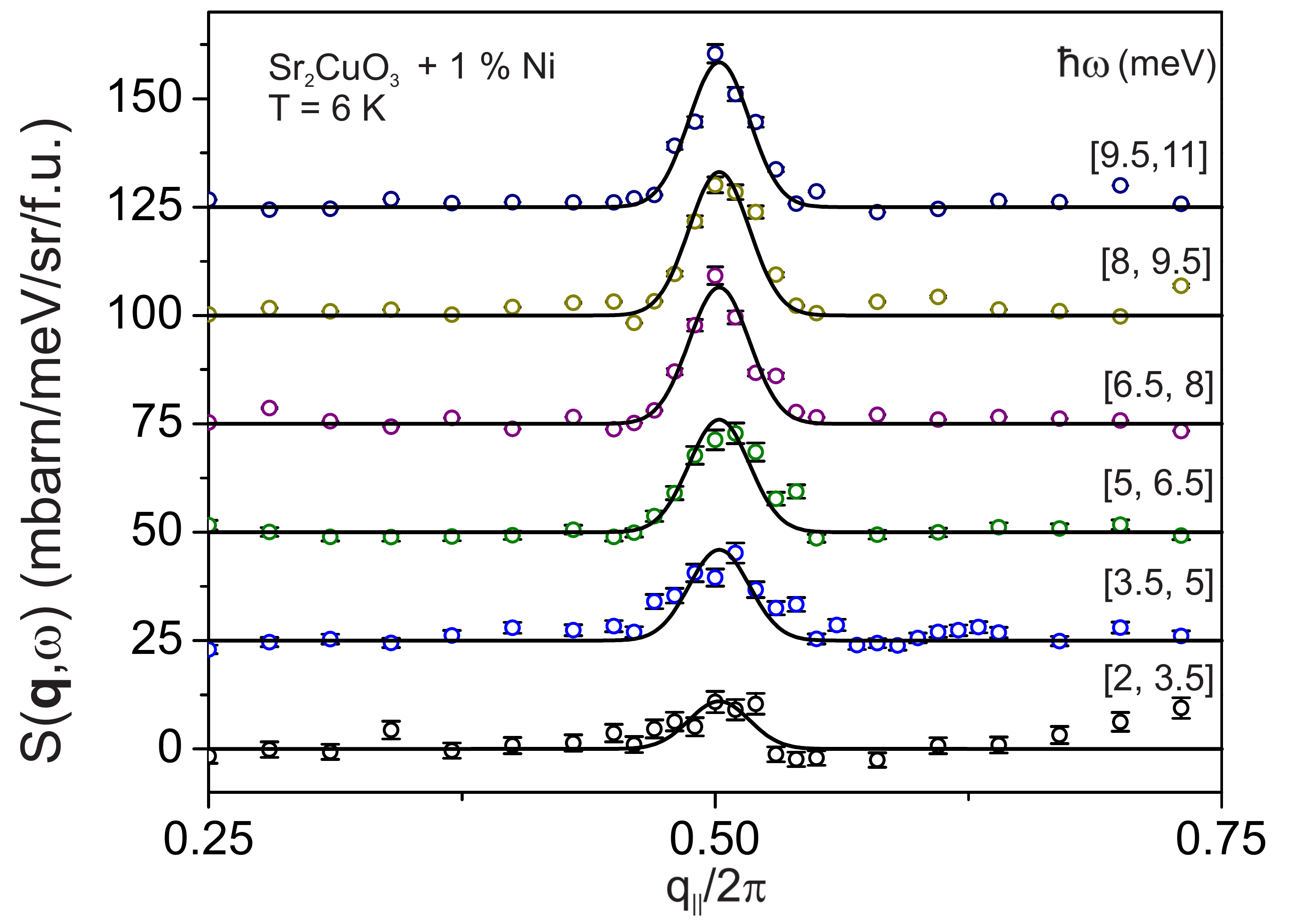}
\caption{Constant-energy cuts of the measured dynamical structure factor in Sr$_2$CuO$_3$ with 1 \% Ni impurities. The data sets are offset by 25 mbarn/meV/sr/f.u. for clarity. The solid lines are Gaussian fits, as described in the text.}
\label{fig:Ni_cut_1}
\end{figure}

\begin{figure}
\centering
\includegraphics[width={\columnwidth}]{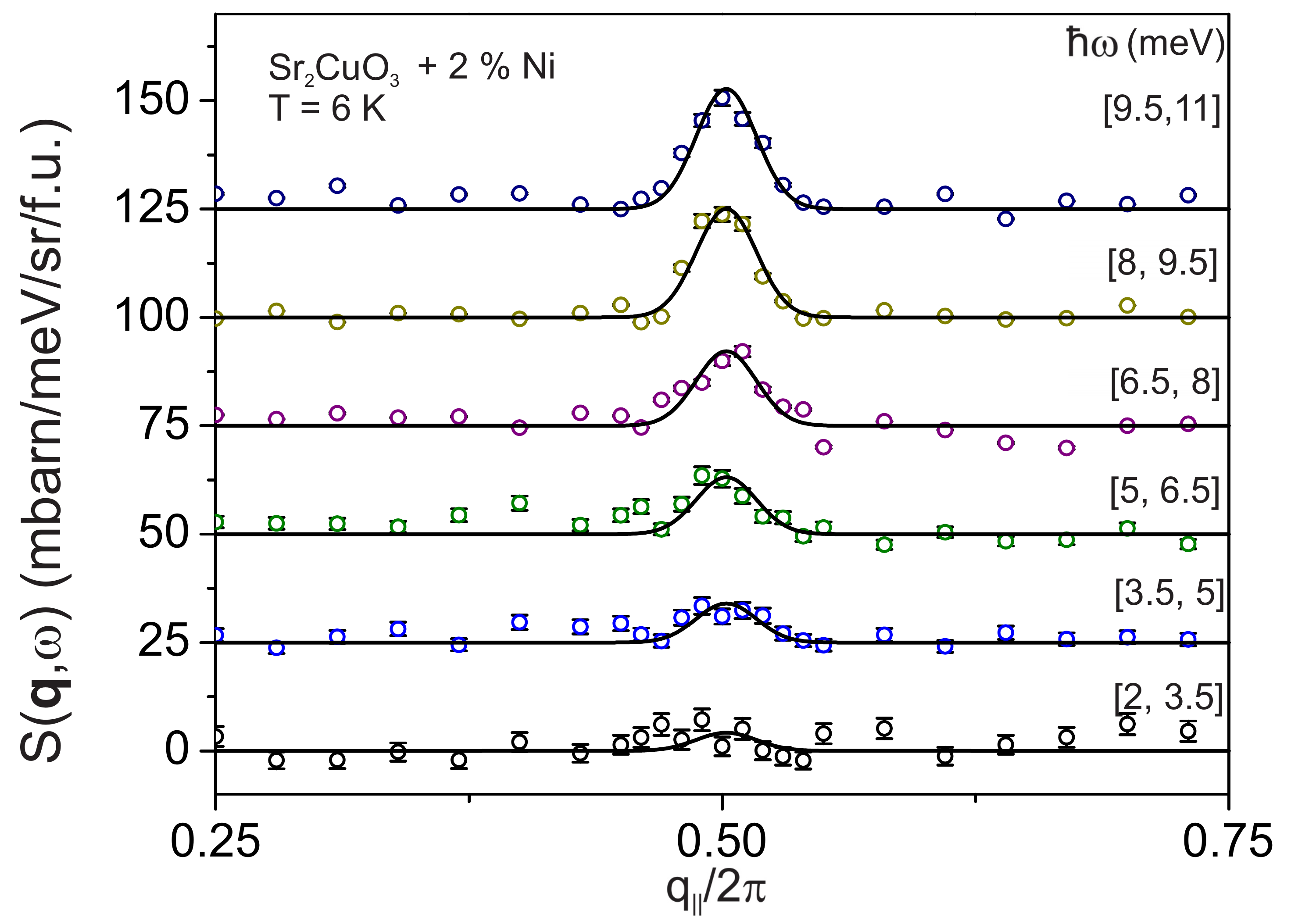}
\caption{Constant-energy cuts of the measured dynamical structure factor in Sr$_2$CuO$_3$ with 2 \% Ni impurities (symbols). The data sets are offset by 25 mbarn/meV/sr/f.u. for clarity. The solid lines are Gaussian fits, as described in the text.}
\label{fig:Ni_cut_2}
\end{figure}

Neutron spectroscopy has been performed at a number of user facilities. The existence of the spin pseudogap in Ni-substituted samples was established using the 4F2 three-axis spectrometer at LLB, using a combination of energy and momentum scans with the final neutron wavevector fixed at k$_f$ =1.55 \AA $^{ −- 1}$. Preliminary measurements on the Ca-substituted compound were performed on the PUMA three-axis spectrometer at FRM-2, using k$_f$ =2.66 \AA $^{ −- 1}$. Most of the measurements were then performed at the SEQUOIA\cite{Granroth2010} time-of-flight spectrometer at ORNL using incident energies of 12, 20 and 50 meV with the high resolution Fermi chopper frequencies set to 240, 300 and 420 Hz respectively. The background was measured by repeating the experiment in the identical configuration with the sample removed and only the sample can in the beampath. This background was subtracted from the data before analysis. The data for the 1 \% Ni-substituted crystal at various temperatures were collected using the IN22 3-axis spectrometer (Collaborative Research Group-CEA at the ILL, Grenoble) with the final neutron wavevector set to k$_f$ =2.66 \AA $^{ −- 1}$. Additional data for samples with Ni impurities were collected using the MERLIN\cite{Bewley2009} spectrometer at ISIS with the high resolution Fermi chopper with a Gd-slit package rotating at 250 Hz and repetition rate multiplication allowing access to incident energies of 10, 20 and 53 meV. In order to achieve quantitative comparison with the theory, the neutron time-of-flight spectra were normalized using a standard vanadium sample. The three-axis measurements at 8 K were then compared with the time-of-flight data at 6 K and a correction factor thus obtained was used to normalize the three-axis data obtained at other temperatures. All the data presented here were obtained at temperature of T = 6 K. The only exception is the measurement of temperature dependence shown in Fig.~\ref{fig:scaling}, where the temperatures are quoted for each dataset.

\section{Experimental Results}

\subsection{Low temperature susceptibility}
To verify the actual concentration of chain-breaking defects in our samples, we performed measurements of their magnetic susceptibility. The results are plotted versus temperature in Fig.~\ref{fig:suscep}. For the  Ni$^{2+}$-substituted samples, impurities are expected to take the place of Cu$^{2+}$ and directly fragment the spin chains. Magnetic susceptibility curves for this scenario have been derived theoretically.\cite{Eggert1992,Fujimoto2004,Sirker2008} The most complete description is given by Sirker et al.\cite{Sirker2008} and allows to quantitatively estimate the actual number of chain breaks. As a characterization tool, this approach has been also validated experimentally.\cite{Karmakar2015, Simutis2016b} By using the same fitting equation as in ref.~\onlinecite{Simutis2016b}, we get very good fits (solid lines in Fig.~\ref{fig:suscep}) to our experimental data. The deviations seen at very low temperatures (note the log scale on the abscissa) can be attributed to weak inter-chain coupling.\cite{Karmakar2015, Simutis2016b}
The fitted chain-break concentration is in an excellent match with the nominal number of impurities, as shown  in the inset of  Fig.~\ref{fig:suscep}.

In the case of Ca-defects, the Ca$^{2+}$ ions are expected to replace Sr$^{2+}$, and not enter the Cu$^{2+}$ chains directly. The magnetic susceptibility measurements in Fig.~\ref{fig:suscep} are consistent with that picture. Even though a large amount of Ca impurities is introduced into the sample, their effect on the magnetic response is minimal. From the observed magnitude of the effect in the 5\% Ca-substituted sample, we can with certainty say that the number of broken links in the Cu-chains is less than 0.6 \%.

\subsection{Spin excitations}

\begin{figure}
\centering
\includegraphics[width={\columnwidth}]{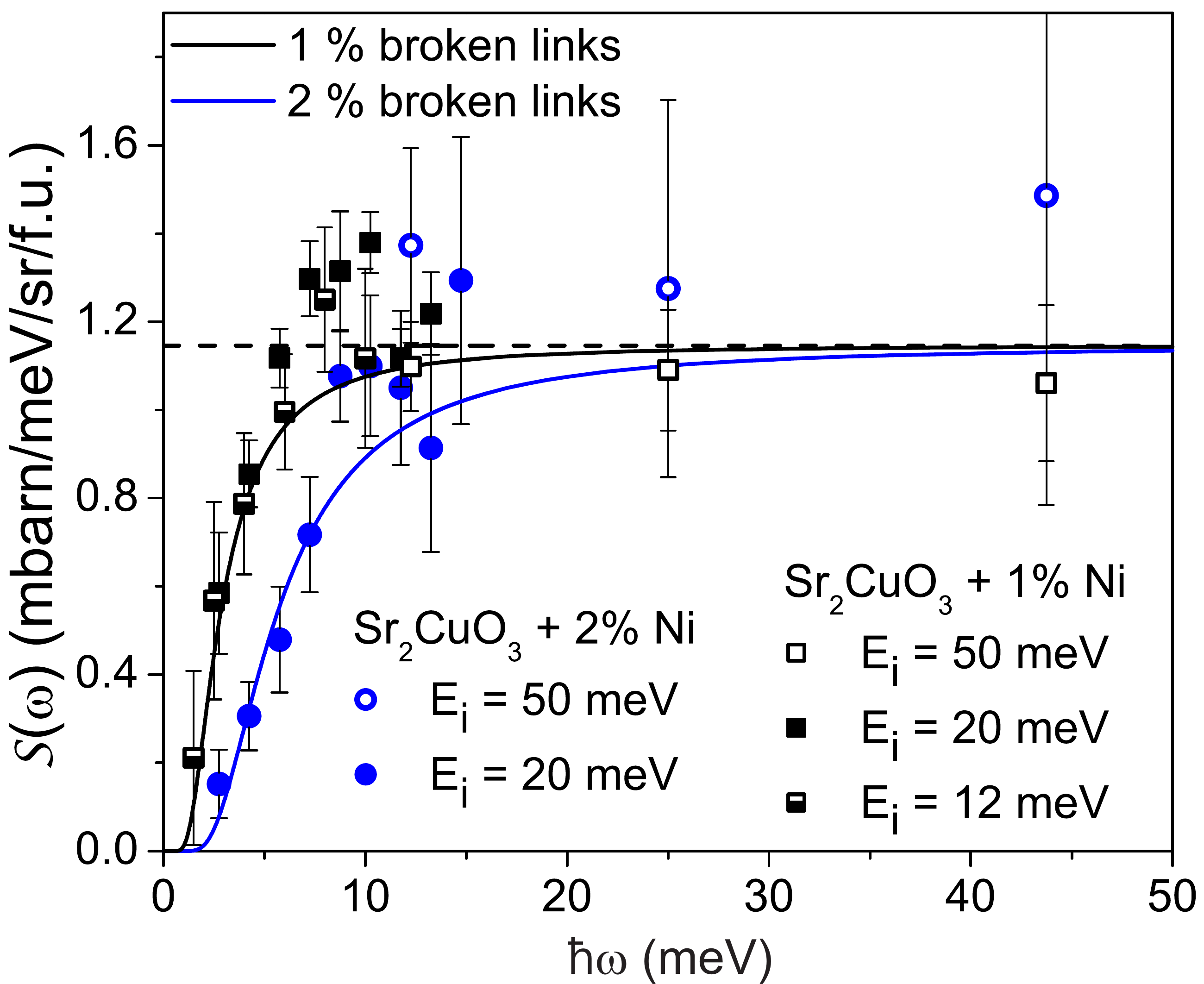}
\caption{Symbols: momentum integrated dynamical structure factor of the 1\% (squares) and 2\% (circles) Ni-substituted samples. The dashed line shows the prediction for an intact chain, whereas the solid lines show the scattering for spin chains with 1\% (black) and 2\% (blue) broken links. \label{fig:Ni_cut}}
\end{figure}

Time-of-flight neutron spectroscopy allows the measurement of the magnetic dynamic structure factor in large portions of the energy-momentum space. When studying one-dimensional systems, the data quality can be improved by integrating over the directions perpendicular to the relevant dimension. In the case of Sr$_2$CuO$_3$, the ratio of inter- and intra-chain exchange constants has been estimated\cite{Yasuda2005} to be $J'/J <10^{-3}$ and therefore such integration can be made use of. A projected slice of such a data set is shown in  Fig.~\ref{fig:Ni_slice} for the Sr$_2$CuO$_3$ with 1 \% and 2 \% Ni impurities. The figure displays the dynamic structure factor as a function of energy transfer $\hbar\omega$ and momentum transfer parallel to the direction of the spin chain $q_{\parallel}$. The vertical rods of intensity correspond to the bottom of the two-spinon continuum. Since the exchange constant in this material is very large, no dispersion can be observed and the width of the observed scattering is due to the resolution of the experimental setup. It is immediately visible that in both cases there is a suppression of low energy states. The scale of this pseudogap is larger in the compound with more impurities.

In order to quantify this effect, the dynamical structure factor was integrated with respect to the momentum transfer along the chain axis. Since the peak width is given by the resolution of the spectrometer, gaussian peaks were used to fit the momentum-cuts, as shown for a selection of cuts in Fig.~\ref{fig:Ni_cut_1} and Fig.~\ref{fig:Ni_cut_2} for 1 \% and 2 \% Ni impurity levels respectively.

At low energy transfers, the intensity of the peaks is low and extracting the correct intensity becomes difficult. Therefore the position and width of the peak were obtained from higher energy cuts, where intensity is substantial, and were then held fixed for the low energy cuts. The thus obtained momentum-integrated (local) dynamical structure factor,
corrected for the anisotropic form factor\cite{Zaliznyak2004} of Cu$^{2+}$, is shown in Fig.~\ref{fig:Ni_cut}. Most of the emphasis was drawn to the low energy part of the spectrum where the pseudogap is observed, and measurements with 50 meV incident energy were performed with shorter counting times. This is especially evident for the high-energy part of the spectrum for 2 \% sample where due to the smaller mass of the sample and shorter counting time the errorbars are substantially larger. Nevertheless, the high energy measurements complement the low energy data set and demonstrate that the dynamical structure factor stays constant at high energy transfers.

As evidenced from Fig.~\ref{fig:Ni_cut}, the system with higher amount of impurities has a more pronounced suppression of low energy states. However, the decrease of intensity is gradual and therefore the extraction of gap size is not straightforward. A simplest way of assigning the value is by quoting the energy where the intensity is halved compared to the expected value. Such an estimate would give values of $\approx$ 3 meV for 1\% Ni impurities and $\approx$ 6 meV for 2 \% Ni impurities. A much better way is to consider the whole lineshape by introducing a model based on the distribution of gap sizes as detailed in the Discussion.

Deviations from the dynamical structure factor of an intact chain were observed at all temperatures. It is best visible, however, at low temperatures, where the momentum-integrated spectrum of an ideal chain is constant in the measured energy range. Therefore most of the discussion is dedicated to the measurements performed at 6 K, where the observed reduction of low energy states is directly due to introduction of impurities.

A series of similar measurements were additionally performed at different temperatures. The results are plotted in Fig.~\ref{fig:scaling}a, together with previous results for 1\%-Ni substituted SrCuO$_2$.\cite{Simutis2013} Here we used scaled coordinates, with $\hbar\omega /2 k_BT$ on the abscissa. The $y$-axis is the momentum-integrated imaginary part of susceptibility multiplied by the corresponding material's exchange constant (J = 241 meV\cite{Walters2009} for Sr$_2$CuO$_3$ and J = 221 meV\cite{Zaliznyak2004} for SrCuO$_2$), for a direct comparison.

\begin{figure}[h!]
\centering
\includegraphics[width={\columnwidth}]{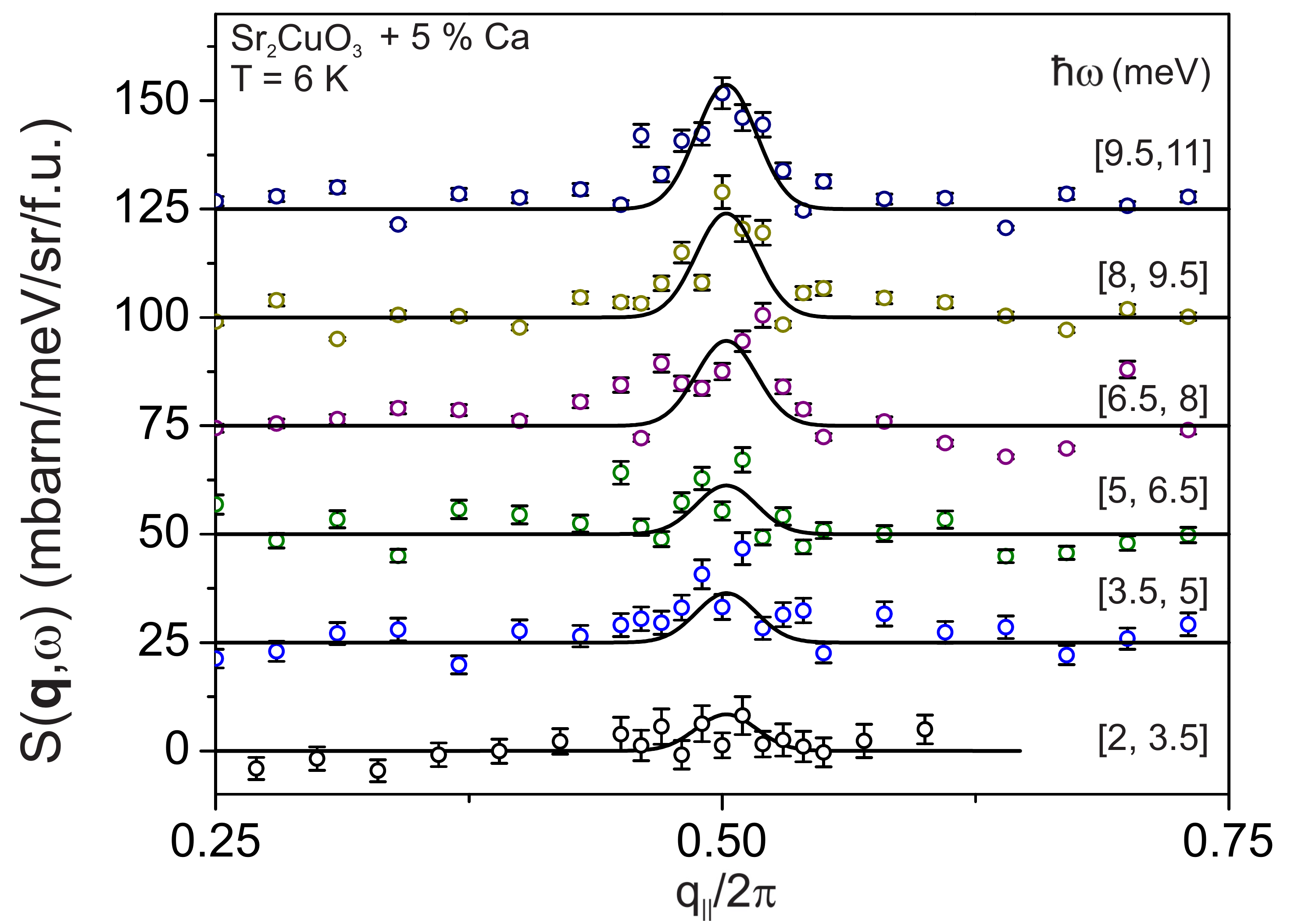}
\caption{Constant-energy cuts of the dynamical structure factor in Sr$_2$CuO$_3$ with 5 \% Ca impurities for selected energy transfers (symbols). The data sets are offset by 25 mbarn/meV/sr/f.u. for clarity. The solid lines are Gaussian fits as described in the text.\label{fig:Ca_doped1}}
\end{figure}

\begin{figure}[h!]
\centering
\includegraphics[width={\columnwidth}]{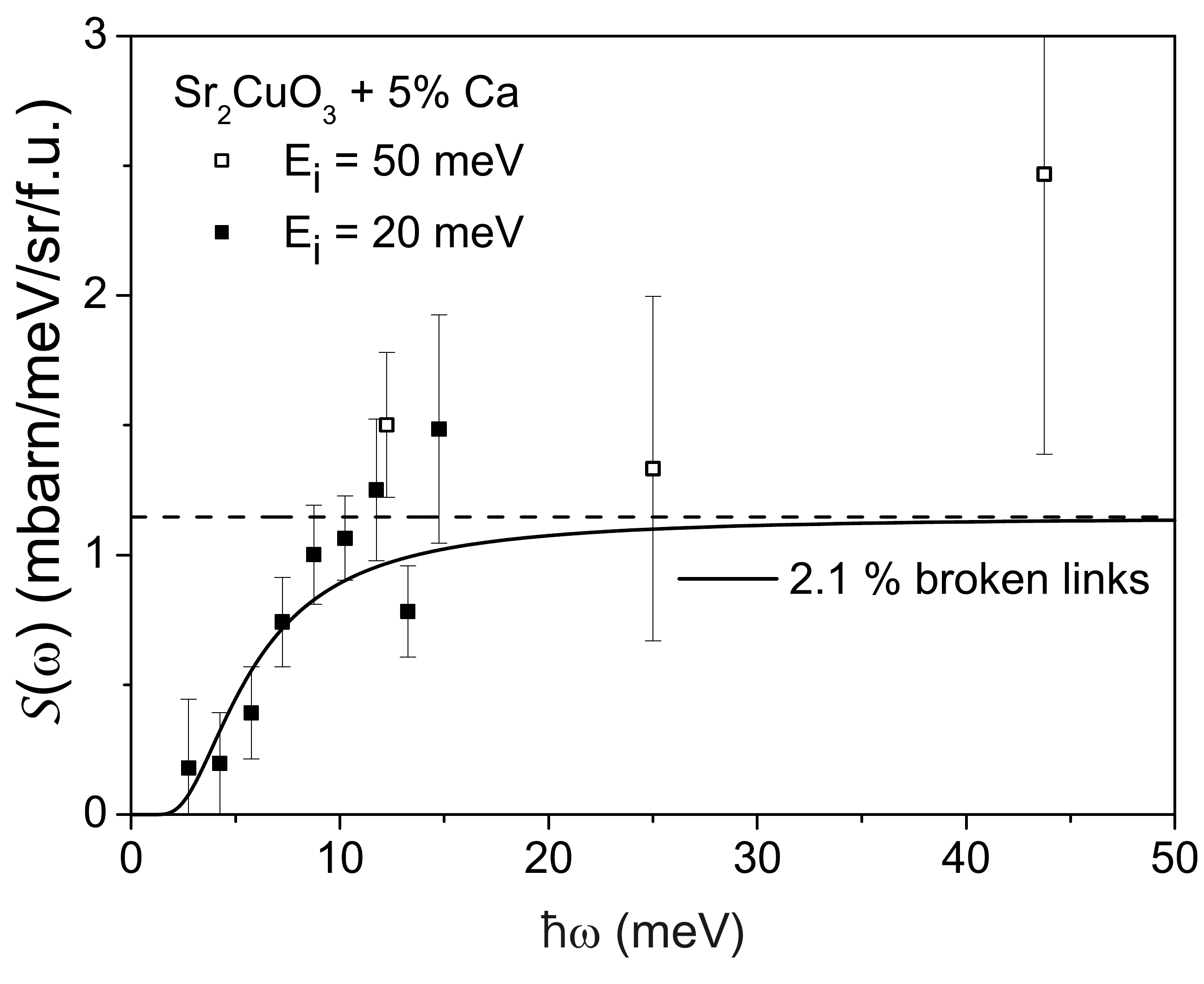}
\caption{Measured low energy excitations in the Ca-substituted Sr$_2$CuO$_3$ in a momentum-integrated form (symbols). The solid line represents a prediction for 2.1 \% broken links.\label{fig:Ca_doped2}}
\end{figure}

The same type of measurements were carried out for the sample with Ca impurities. The data, while of poorer quality due to a smaller sample, show a clear suppression of low energy states (Fig.~\ref{fig:Ca_doped1}). Its value can be estimated from the momentum-integrated dynamical structure factor shown in Fig.~\ref{fig:Ca_doped2}. Somewhat surprisingly, it is similar in magnitude to the pseudogaps observed in the samples with 1\%- and 2\%-Ni impurities.

\section{Discussion}

\subsection{Pseudogap and impurity concentration}

The data shown in Fig.~\ref{fig:Ni_slice} are a vivid illustration of the pseudogap dependence on impurity concentration.
As argued earlier,\cite{Simutis2013} the pseudogap arises from  a confined motion of spinons in chain fragments bound by defect sites. A finite length $L$ of such fragments ensures a gapping and overall discretization of the excitation spectrum with the spacing of $\Delta = 3.65 \times J / L$, where $J$ is the intrachain exchange constant.\cite{Eggert1992} The observed spectrum is defined by the statistical length distribution of fragments in which the quasiparticles are trapped. The resulting dynamic structure factor can be factored into that of a defect-free chain and a defect-related envelope function.\cite{Simutis2013} The envelope function depends on the defect concentration $x$ and can be expressed as:\cite{Simutis2013}

\begin{equation}
F(\omega) =\left(\frac{\Delta_0 \times x}{2\hbar\omega}\right)^2\sinh^{-2}\left(\frac{\Delta_0 \times x}{2\hbar\omega}\right) \mathrm{,}
\end{equation}

where $\Delta_0 = 3.65 \times J$.

The envelope function effectively describes the modification of spin chain spectrum in a bulk material due to randomly distributed impurities. Different distributions would lead to different shapes of the envelope function.

Since the magnetic dynamical structure factor of a defect-free chain is known in absolute units, and the envelope function is uniquely expressed through the exchange constant and the defect concentration, the model has {\it no adjustable parameters}. Nevertheless, as shown in solid lines in Fig.~\ref{fig:Ni_cut} it provides an excellent description of our data {\it on the absolute scale}. It is worthwhile noting that there seems to be a deviation of the data points from the prediction at energy transfers around 10 meV. While the origin of this deviation is unclear, there are two possibilities for such behavior. First, it is conceivable that our assumption of a random distribution of the defects is too simplistic. A reduced probability of very short chain segments would lead to the observed spectrum with an earlier saturation. Second, this deviation could be due to an additional peak which may warrant further attention from the theoretical point of view.

\begin{figure}[h!]
\centering
\includegraphics[width={\columnwidth}]{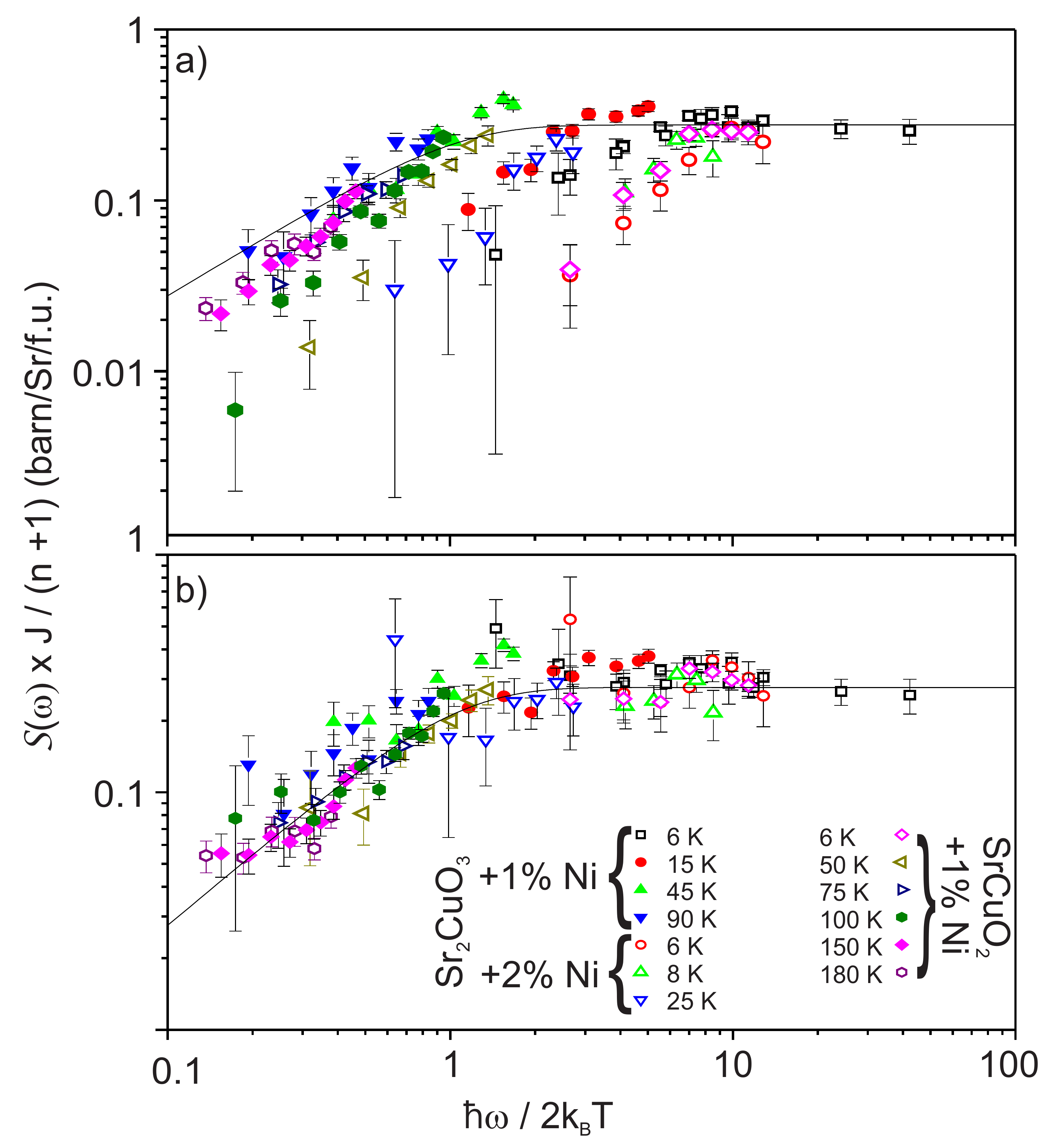}
\caption{The dynamical structure factor expressed in a universal scaling form before (a) and after (b) correction by the pseudogap function. The empty(full) symbols correspond to time-of-flight(three-axis) measurements. The solid line is the theoretical prediction for the S = 1/2 chain. While the raw data in (a) display the breakdown of scaling behavior, it can be restored by taking into account the T-independent envelope function. A corrected dataset in (b) shows scaling at all studied temperatures. \label{fig:scaling}}
\end{figure}

As pointed out in Ref.~\onlinecite{Simutis2013}, the envelope function is independent of temperature. For this reason, the model is expected to work even at elevated temperatures. In particular, after a normalization by the envelope function, finite-temperature data are expected to obey the scaling relations for spin correlations in defect free chains.  In the latter, the imaginary part of local susceptibility $\chi''(\omega)=S(\omega)/[n(\omega)+1]$, is a universal function of $\omega/T$.\cite{Schulz1986,Dender1997,Lake2005} The scaling function is known,\cite{Schulz1986,Dender1997} and is plotted as a solid line in Fig.~\ref{fig:scaling}. The raw data for our Ni-substituted samples clearly violate scaling. However, when normalized by the respective envelope function, they produce a convincing data collapse across all materials, temperatures and defect concentrations.

Additional confidence in such interpretation of the pseudogap is provided by the recent NMR experiments of 1\% and 2 \% Ni-substituted Sr$_2$CuO$_3$, where the characteristic temperature of the drop of spin-lattice relaxation rate was found to scale with the impurity concentration.\cite{Utz2015} Since the relaxation mechanism in such an experiment is due to scattering of thermally excited spinons, the observation of Ref. \onlinecite{Utz2015} also implies scaling of pseudogap with impurity concentration. We believe that such scaling of pseudogap with impurity concentration should persist up to higher concentrations until clustering of impurities becomes relevant. Experimentally, it would become more challenging to observe it, since at higher energies, the phonons start to obscure the neutron scattering spectrum and NMR measurements would require a high-temperature apparatus.

\subsection{Defects vs. scattering centers}

The spectrum measured in the  5\% Ca-substituted compound at a first glance seems to be at odds with this picture. As mentioned above, off-chain Ca$^{2+}$ defects have virtually no effect on the magnetic susceptibility. At the same time, neutron scattering clearly shows a pseudogap (Figs.~\ref{fig:Ca_doped1} and Fig.~\ref{fig:Ca_doped2}). Using the previously proposed model, and treating the number of chain-breaks as a fitting parameter for the neutron data we deduce as much as 2.1(3)\% chain-defects, well above the $<0.6$\% susceptibility estimate. This apparent contradiction is resolved if we recall that contributions to magnetic susceptibility and the spectral pseudogap are of different origin. The former is caused by end-chain degrees of freedom that form in finite-length chain fragments.\cite{Sirker2008} In contrast, the latter is due to a confined motion of spinons. Any disturbances that scatter spinons will contribute to the pseudogap, but only true chain-severing defects will add to low-temperature susceptibility. In this picture, Ca$^{2+}$ ions are responsible for locally distorting the crystal lattice, and thereby inducing weak scattering centers in the chains with no effect on the isothermal magnetic response. A similar mechanism may account for the  almost 2-fold enhancement of the ``effective''defect concentration observed in Ni-substituted SrCuO$_2$.\cite{Simutis2013} Due to the double-chain structure of that compound, each Ni$^{2+}$ center severs one spin chain, but additionally induces a scattering center in the adjacent chain.

\section{Conclusion}

In summary, we have performed inelastic neutron scattering experiments which demonstrate the existence of a spin pseudogap in Sr$_2$CuO$_3$ with Ni- and Ca-impurities. The origin of the pseudogap is found to be due to scattering centers induced by the defects with the magnitude scaling proportionately to the concentration of impurities. The proposed model explains the measured data in a broad temperature range.

\section{Acknowledgements}
This work was supported by the Swiss National Science Foundation, Division 2. The work at IFW has been supported by the Deutsche Forschungsgemeinschaft through the D-A-CH Project No. HE 3439/12 and the European Commission through the LOTHERM project (Project No. PITN-GA-2009-238475). The neutron scattering experiments at Oak Ridge National Laboratory’s Spallation Neutron Source were sponsored by the Scientific User Facilities Division, Office of Basic Energy Sciences, U.S. Department of Energy. The experiment at the MLZ was financially supported by the EU Framework 7 programme NMI3. Additional support for the work at ILL was provided by the Swiss State Secretariat for Education, Research and Innovation (SERI) through a CRG-grant.

\bibliography{S2CO3neutronbib}

\end{document}